\documentclass[fleqn,twocolumn,aps,prl,superscriptaddress,showpacs]{revtex4}
\usepackage{amsmath,bm}%
\usepackage{graphicx}%

\usepackage{CJK}
\begin{document}

\begin{CJK*}{GBK}{song}

\title{Mach-like emission from nucleon scattering in proton-nucleus reaction}
\author{G. Q. Zhang}
\affiliation{Shanghai Institute of Applied Physics, Chinese
Academy of Sciences, Shanghai 201800, China} \affiliation{Graduate
School of the Chinese Academy of Sciences, Beijing 100080, China}
\author{Y. G. Ma \footnote{Corresponding author. E-mail address: ygma@sinap.ac.cn }}
\affiliation{Shanghai Institute of Applied Physics, Chinese
Academy of Sciences, Shanghai 201800, China}
\author{X. G. Cao}
\affiliation{Institute of Modern Physics, Chinese Academy of
Sciences, Lanzhou 730000, China}
\author{C. L. Zhou}
\affiliation{Shanghai Institute of Applied Physics, Chinese
Academy of Sciences, Shanghai 201800, China} \affiliation{Graduate
School of the Chinese Academy of Sciences, Beijing 100080, China}

\date{\today}

\begin{abstract}
The fast-stage nucleon emission of proton-nucleus (pA) reactions
from 300$A$ MeV to 1.8$A$ GeV has been investigated by the quantum
molecular dynamics model. It is found that the sideward angular
spectrum of nucleon emission presents an interesting Mach-like
structure  at the early stage of the collision (tens of fm/c). The
sideward angular peak value varies from about 45$^\circ$ to near
73$^\circ$, depending on the bombarding energy. Nucleons emitted
from the vicinity of the sideward peak tend to have a fixed
momentum value about 0.5 GeV/c, independent of the bombarding
energy as well as the impact parameter. Additionally, the sideward
angular peak value is almost independent of equation of state,
indicating that binary collision at the early fast stage in the
intermediate energy pA reaction plays an important role for
emergence of Mach-like emission.
\end{abstract}
\pacs{25.40.-h}

\maketitle

\section{\label{sec:level1}Introduction}

Canonical emission is of very interesting phenomenon occurs in
different fields in physics. Recently BNL Relativistic Heavy Ion
Collider data have shown that a hot and dense quark-gluon plasma
(QGP) medium is created in ultrarelativistic heavy-ion collision
(HIC). The QGP behaves like an almost perfect fluid and to be
opaque to jets created in the initial stage of the collision. The
experimental  dihadron correlation function
\cite{STAR1,STAR2,Phenix1,Phenix2} exhibits an interesting
double-peak structure at angles opposite to the trigger jet. It
has been suggested \cite{Stocker,Cas} that such a structure could
be an evidence for Mach cone. However, different theoretical
interpretations have been provided, such as Cherenkov-like gluon
radiation model~\cite{Koch}, shock wave model in hydrodynamic
equations~\cite{Cas},  jet deflection~\cite{deflection} and strong
parton cascade mechanism \cite{MaGL} etc. Since it is still
unclear that the main mechanism for the emergence of the double
peak structure, many experimental and theoretical works suggest
that it should depend on the nature of the hot and dense matter
created in the collisions. Very recently, Betz et al. suggested
that the conical emission  can also arise due to averaging over
many jet events in a transversally expanding background in
ultrarelativistic heavy-ion collisions. Furthermore, they found
that the apparent width of the away-side shoulder correlation is
insensitive to the details of the energy momentum deposition
mechanism as well as to the system
size~\cite{Gyulassy-flow-driven}. Even though much effort on
dihadron correlation has been done, the mechanism of such double
peak structure is still in debating.

On the other hand, Mach-like structure has been noticed in heavy
ion collision at several GeV energy in earlier time. For an
example, Aichelin et al. had used quantum molecular dynamics (QMD)
model to study the asymmetry reaction system $^{20}$Ne +
$^{197}$Au at 1050$A$ MeV in central collisions
\cite{Aichelin1988}. A shock wave picture was depicted at the fast
stage. While the projectile nucleons punch through the heavy
target with a supersonic speed, a strong compression about
2$\rho_0$ can be reached and a strong transverse force is put on
the projectile surface. Thus, nucleons on the projectile surface
will get a transverse velocity and emit in the sideward direction.
The picture is very similar to Mach Cone phenomenon. Recently, Rau
et al. \cite{Rau2010} adopted the hydrodynamic model
\cite{Rischke1995} and UrQMD \cite{Bass1998} to simulate the
Mach-like wave in the asymmetric system $^{20}$Ne + $^{238}$U at
1-20$A$ GeV. They got clear Mach cone structure  as the picture
emerged in hydrodynamic dynamics. While, in the default UrQMD
dynamics, they also got similar sideward peak excitation function,
although UrQMD has a different mechanism forming the sideward peak
and gets a broader width in the sideward angular distribution.

Considering that  proton-nucleus reaction (pA) is relative simple
in reaction mechanism in comparison with the heavy ion reactions,
there is a potential advantage to give some hints to understand
the mechanism of Mach-like phenomenon in HIC by investigating the
fast-stage nucleon emission.  In past several decades, the
intermediate energy proton-induced reactions (pA)  play important
roles in the wide applications and fundamental research fields
\cite{Kowalczyk2008,Herbach2006,Ou2007}. However, the mechanism of
pA collision is still not well understood, especially on how to
form the sideward peak angular distributions of intermediate mass
fragments (IMF) and light charged particles (LCP)
\cite{Remsberg1975,Hirata2002,Hsi1998,Hsi1999,Ricciardi2006}. In
early 1970s, Remsberg and Perry \cite{Remsberg1975} showed that
the angular distribution of IMF dominates 60$^\circ$ to 70$^\circ$
which is coincided with the angular distribution of Mach Shock
conical emission predicted by hydrodynamics models. Hirata et al.
\cite{Hirata2002} used a newly developed non-equilibrium
percolation (NEP) model and concluded that a doughnut shape
structure results in the IMF sideward peak angular distribution.
Hsi et al. \cite{Hsi1998,Hsi1999,Ricciardi2006} performed
exclusive experimental studies on pA reactions and claimed that
the sideward peak of IMF originates in kinematic-focusing effects
associated with statistical and thermal multifragmentation of an
expanding residue. They argued that the sideward peak was not
likely to be the result of shock wave, based on their experimental
inclusive angular distribution without sideward peak for LCPs.
Using angular correlation analysis, they also denied the sideward
peak of IMF angular distribution could come from the breakup of
exotic geometric shapes.

Although different physical hypotheses are employed, Quantum
Molecular Dynamics model could demonstrate similar phenomenon that
hydrodynamical model predicts. In this Letter, we adopt quantum
molecular dynamics model to investigate pA reactions. The reaction
system p + $^{208}$Pb is taken as an example. Various energies and
impact parameters, as well as hard or soft EOS, are scanned
systematically. The focus is concentrated on nucleon emission at
the fast stage of the collision, including the kinetic energy
distribution and polar angular distribution.

\section{\label{sec:level2} Isospin Dependent Quantum Molecular Dynamics model}

Quantum Molecular Dynamics model bases on an n-body theory, which
simulates heavy ion reactions at intermediate energies on an event
by event basis \cite{Hartnack1998,Aichelin1991}. The Isospin
Dependent QMD (IDQMD) is an extension version of QMD, which is
suitable to describe from Fermi energy up to 2$A$ GeV with the
isospin effects considered: different density distribution for
neutron and proton, the asymmetry potential term in mean field,
different experimental cross-section for neutron-proton (np) and
proton-proton (pp,nn), Pauli blocking for neutron and proton
separately \cite{Ma2006a,Cao2010}. Each nucleon is presented in a
Wigner distribution function with a width $\sqrt{L}$ (here $L$ =
2.16 ${\rm fm}^2$) centered around the mean position
$\vec{r_i}(t)$ and the mean momentum $\vec{p_i}(t)$, $
\phi_i(\vec{r},t) = \frac{1}{{(2\pi L)}^{3/4}}
exp[-\frac{{(\vec{r}- \vec{r_i}(t))}^2}{4L}] exp[-\frac{i\vec{r}
\cdot \vec{p_i}(t)}{\hbar}]. $ The mean field in IDQMD model is: $
U(\rho) = U^{\rm Sky} + U^{\rm Coul}  + U^{\rm Yuk} + U^{\rm sym}
$, where $U^{\rm Sky}$, $U^{\rm Coul}$, $U^{\rm Yuk}$, and $U^{\rm
sym}$  represents the Skyrme potential, the Coulomb potential, the
Yukawa potential and the symmetry potential interaction,
respectively \cite{Aichelin1991}. The Skyrme potential is: $
U^{\rm Sky} = \alpha(\rho/\rho_{0}) +
\beta{(\rho/\rho_{0})}^{\gamma} $, where $\rho_{0}$ = 0.16
{fm}$^{-3}$ and $\rho$ is the nuclear density. The parameters
$\alpha =-356$ MeV, $\beta =303$ MeV, and $\gamma = 7/6$,
correspond to a soft EOS, and $\alpha = -124$ MeV, $\beta = 70.5$
MeV, and $\gamma = 2$, correspond to a hard EOS. $U^{\rm Yuk}$ is
a long-range interaction (surface) potential, and takes the
following form: $ U^{Yuk} = ({V_y}/{2}) \sum_{i \neq
j}{exp(Lm^2)}/{r_{ij}}
\cdot[exp(mr_{ij})erfc(\sqrt{L}m-{r_{ij}}/{\sqrt{4L}})
-exp(mr_{ij})erfc(\sqrt{L}m+{r_{ij}}/{\sqrt{4L}})] $ with $V_y =
0.0074$GeV, $m = 1.25{fm}^{-1}$,  $L = 2.16$ fm$^{2}$, and
$r_{ij}$ is the relative distance between two nucleons. The
strength of symmetry potential is $C_{sym}$ = 32 MeV.

IDQMD treats the many body state explicitly, and contains
correlation effects to all orders and deals with fragmentation and
fluctuation of HICs. To recognize fragments produced in HICs, a
simple coalescence rule is used with the criteria $\Delta$r = 3.5
fm and $\Delta$p = 300 MeV/c between two considered nucleons.
Thus, nucleons dominated in Fermi motion will be limited in the
target.

\section{\label{sec:level3}Results and Discussions}

\begin{figure}
\includegraphics[scale=0.42]{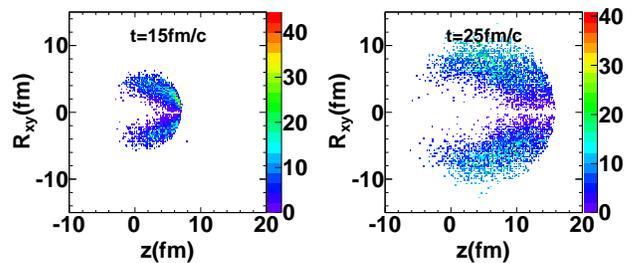}
\vspace{-0.1truein} \caption{\footnotesize (Color online) Position
correlation 2D histogram $R_{xy}$ versus $z$ in a cylindrical
coordinate system. The reaction system is 1$A$ GeV p+ $^{208}$Pb
at
 $b$ = 1 fm with the hard EOS. Left: Evolution time is 15 fm/c.
Right: Evolution time is 25 fm/c. A linear scale plot is used.
}\label{Fig1}
\end{figure}

Conical emission  structure at the early stage of pA reaction has
been observed in our IDQMD calculation. As Fig.\ref{Fig1} presents
for p + Pb at 1$A$ GeV, Mach-like structure develops  at 15 fm/c,
while the head of the conical structure disappears at 25 fm/c.

\begin{figure}
\includegraphics[scale=0.42]{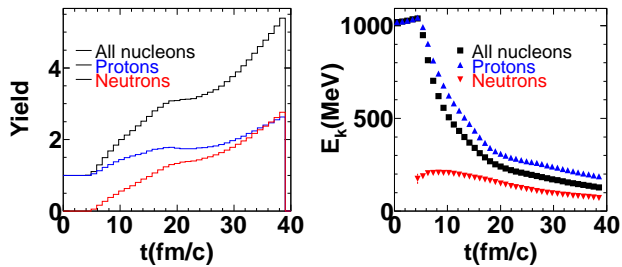}
\vspace{-0.1truein} \caption{\footnotesize (Color online)Left: The
yield of nucleons at different time.
Right: The average kinetic energy at different time for all
nucleons (black square), protons (blue up triangle) and neutrons
(red down triangle).  The same reaction system and condition as
Fig. 1.  }\label{Fig2}
\end{figure}

Fig.\ref{Fig2} shows that the yield of nucleons and the average
kinetic energy of nucleons evolve with time. The yield shows a
stable increase for both protons and neutrons. However, at 15
fm/c, including the protons emitted from projectile, there are
only 2.5 nucleons at each event on average. This means that at
each event, there exists no Mach-like structure, because, at
least, three nucleons are required to form such structure. In
addition, the average kinetic energy of nucleons presents a
dropping situation. This, therefore, gives a hint that fast
nucleons emit at the early stage and slow nucleons emit at the
latter stage. The trend that neutron yields faster than proton, is
consistent with the phenomenon which Ma et al. has found
\cite{Ma2006a}.

\begin{figure}
\includegraphics[scale=0.4]{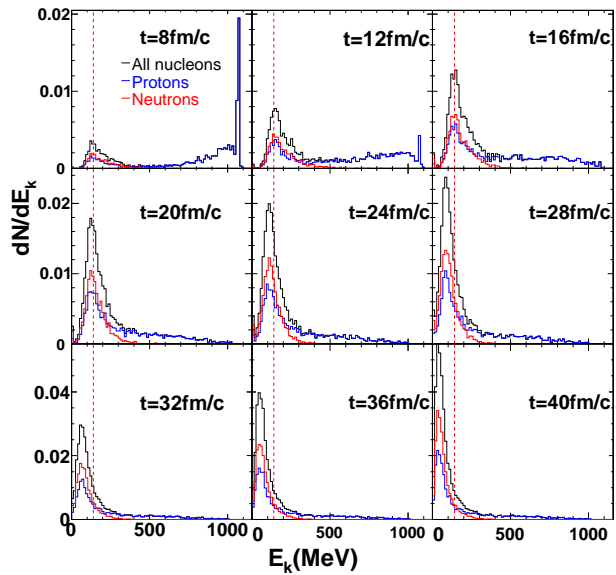}
\vspace{-0.1truein} \caption{\footnotesize (Color online):
Evolution of the kinetic energy distribution of nucleons with
time.
 A fixed peak can be found near $E_k$ = 140 MeV before 20 fm/c, then the low
energy part comes out and the former 140 MeV peak is submerged.
The same reaction system and condition as Fig. 1. }\label{Fig3}
\end{figure}

\begin{figure}
\includegraphics[scale=0.4]{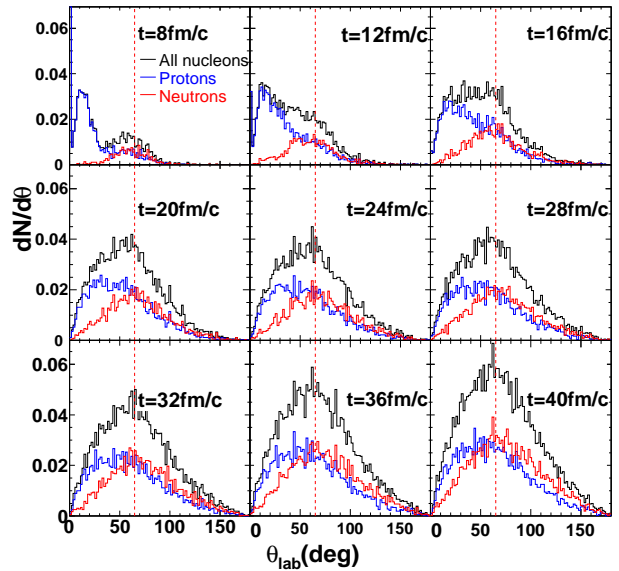}
\vspace{-0.1truein} \caption{\footnotesize (Color online):
Evolution of the polar angle distribution of nucleons with time. A
fixed sideward peak can be found near 65$^\circ$ in laboratory
system.  The same reaction system and condition as Fig.
1.}\label{Fig4}
\end{figure}

To show the dynamics process in detail, we select several time
points during the reaction evolution to see the kinetic energy
($E_k$) spectra (Fig.~\ref{Fig3}) and polar angular distribution
(Fig.~\ref{Fig4}). For all nucleons,  a roughly fixed $E_k$ peak
near 140 MeV can be found before 20 fm/c, then $E_k$ smears to
lower energy. For protons, there exists a high energy peak at the
early stage, and later on it turns into a high energy tail which
stems from those induced protons, distorted only a little by the
mean field.

Hsi et al.\cite{Hsi1999} had measured the polar angular
distribution for \textquotedblleft gray proton\textquotedblright
(100MeV $< Ep < $400MeV) and found no sideward peak. Basing on
this consideration, they denied the possibility of
shock-wave-like effects in pA reaction. However, it is not so easy
to select suitable kinetic energy cut to separate the nucleons
emitting at the early fast stage from the ones evaporating at the
latter stage. In addition, the efficiency for detecting high
energy nucleons should be carefully considered.

In pA reaction, the maximal density is just a little more than
normal nuclear matter density $\rho_0$. Therefore it is hard to
explain the transverse emission by the way the Ref.
\cite{Aichelin1988} does, which needs high compression gradient.
In IDQMD, with the reaction process going on, the small value of
angular peak, composed by mostly high energy induced protons, gets
more and more feeble, while a fixed sideward peak shows up at
about 65$^\circ$ for all nucleons, then a thermal part comes out
with a peak at $\sim$90$^\circ$, which will dominate at the latter
isotropic stage ( see Fig.\ref{Fig4}). The incident protons would
also get the chance of collision with other nucleons in the
target. The chance of collision is determined by experimental
proton-proton and proton-neutron cross-section, including the
elastic and the inelastic channels. After the binary collision,
the projectile proton shares its kinetic energy and momentum with
its collision partner, which results in the sideward angular
distribution peak at the early stage of pA reaction.

\begin{figure}
\includegraphics[scale=0.42]{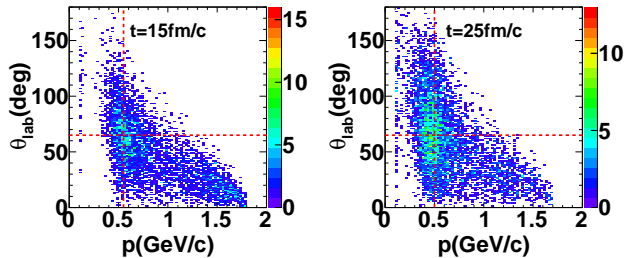}
\vspace{-0.1truein} \caption{\footnotesize (Color online):
Correlation between the momentum and polar angular of nucleons at
15fm/c (left) and 25fm/c (right).  Nucleons near 65$^\circ$
sideward peak have a fixed momentum value about 0.53 GeV/c
(140MeV) in Laboratory system.  The same reaction system and
condition as Fig. 1. }\label{Fig5}
\end{figure}

The correlation between the momentum and polar angular of nucleons
is also studied. Nucleons coming from the the vicinity of the
angular peak tend to get a fixed momentum value 0.53GeV (~140MeV)
at the early stage (15fm/c). At the following stage (25fm/c), the
sideward emission nucleons move their momentum peak to lower value
and extend their angular distribution width at the same time
(Figure \ref{Fig5}). The situation is very similar to Mach shock
structure that ideal hydrodynamical models have predicted.

Additionally, different impact parameters, energies and EOS are
investigated systematically. Nucleons emitting within momentum
range 0.3GeV/c $<$ P $<$ 0.6GeV/c are selected in the following
study. After the head of projectile protons punches through the
center of $^{208}$Pb target, five angular peak values are sampled
continuously, with a time step 1 fm/c. The average angular peak
value is then calculated from these five peak values as the final
result. Excitation function of the sideward angular peak value for
p + $^{208}$Pb is presented for the hard EOS (Fig. 6). The
sideward angular peak values increase with the beam energy from
about 45$^\circ$ at 330MeV up to and a limited 73$^\circ$ at
1.81GeV. Soft EOS gives the same values of angular peak values and
bombarding energy dependence as the hard EOS case, and also shows
insensitivity to impact parameter (not shown here due to the
limited space). Overall, the results are independent of the impact
parameter and EOS.

\begin{figure}
\includegraphics[scale=0.42]{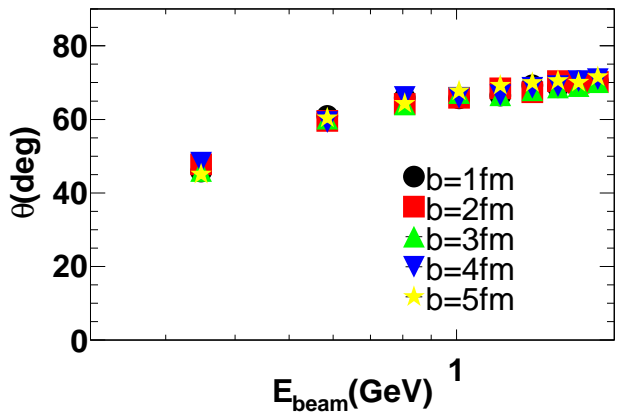}
\vspace{-0.1truein} \caption{\footnotesize (Color online):
Excitation function of the sideward angular peak value of
nucleons. The system is p + $^{208}$Pb from energy 330$A$ MeV to
1.81$A$ GeV with impact parameter from $b$ = 1 fm to $b$ = 5 fm
and hard EOS. The error bars are smaller than the marker-size.
}\label{Fig6}
\end{figure}

\section{\label{sec:level4}Conclusion and summary}

In summary, we apply quantum molecular dynamics model to revisit
the early stage nucleon emission in pA reaction. It is found, for
the first time, that the peak of sideward angular distribution for
the early emitting nucleons is energy dependent. With the
increasing of  beam energy, the sideward angular peak value
increases to a limit value 73$^\circ$, while the momentum of the
emitting protons tend to get a fixed value 0.53 GeV/c (140 MeV).
However, the  peak of sideward angular distributions are
independent of impact parameters and EOS. This Mach-like
phenomenon is similar to the result that hydrodynamical models
have predicted, although different physical interpretation are
employed. The binary nucleon collision plays the essential role,
which results in the sideward angular emission. In event-level,
there are only 2-3 nucleons emitting at the early stage (before 20
fm/c), so there exists no Mach-like structure in the event basis.
In this case, we are unable to apply the angular correlation
method to extract the Mach-like structure information event by
event. In the present model, the Mach-like structure can be
regarded as the scattering effect mostly by the bombarding proton
with the target nucleons summing over each event, which is very
different from bulk collective hydrodynamical behavior. This
mechanism may also give some hints to the Mach-like cone structure
in ultra-relativistic energy
\cite{Cas,Stocker,MaGL,Gyulassy-flow-driven,deflection,Han}. As
far as the experiment is concerned, the key is to separate the
nucleons emitting at the early stage from those coming from the
later stage and to improve the detecting efficiency of the forward
and sideward emitting nucleons, especially for those
\textquotedblleft grey nucleons\textquotedblright\cite{Hsi1999}.

This work was supported in part by  the National Natural Science
Foundation of China under Grant No. 11035009, 10979074, 10905085,
10875159, and the Shanghai Development Foundation for Science and
Technology under contract No. 09JC1416800, and the Knowledge
Innovation Project of the Chinese Academy of Sciences under Grant
No. KJCX2-EW-N01.

\end{CJK*}
\end{document}